\documentclass[twocolumn,floatfix,superscriptaddress,aps,prr]{revtex4-1}
\usepackage{graphicx}
\usepackage{amsmath,mathtools}
\usepackage{braket}
\usepackage{bm}
\usepackage{booktabs}
\usepackage{hyperref}
\usepackage{pgfplots}
\usepackage{multirow}
\usepackage{amsfonts}
\usepackage{calc}
\usepackage{siunitx}
\usepackage{array}
\usepackage{color}

\usepackage{tikz}
\usepackage[utf8]{inputenc}
\usepackage{pgfplots}
\usepgfplotslibrary{groupplots,dateplot}
\usetikzlibrary{patterns,shapes.arrows}
\pgfplotsset{compat=newest}

\definecolor{ashgrey}{rgb}{0.7, 0.75, 0.71}
\definecolor{aurometalsaurus}{rgb}{0.43, 0.5, 0.5}

\hypersetup{
    colorlinks,
    linkcolor={red!50!black},
    citecolor={blue!50!black},
    urlcolor={blue!80!black}
}

\AtBeginDocument{
\heavyrulewidth=.08em
\lightrulewidth=.05em
\cmidrulewidth=.03em
\belowrulesep=.65ex
\belowbottomsep=0pt
\aboverulesep=.4ex
\abovetopsep=0pt
\cmidrulesep=\doublerulesep
\cmidrulekern=.5em
\defaultaddspace=.5em
}

\begin{document}

\title{Quantum Induced Broadening- A Challenge For Cosmic Neutrino
Background Discovery}

\author{Shmuel Nussinov}
\affiliation{School of Physics and Astronomy, Tel Aviv University,
6997801 Tel Aviv, Israel}

\author{Zohar Nussinov}
\affiliation{Department of Physics, Washington University, St.\ Louis,
MO 63130, USA}

\date{\today}

\begin{abstract}
A recent preprint by Cheipesh {\it et al.} pointed out that the
zero-point motion of Tritium atoms bound to Graphene may blur the
measured energies of $\beta$ electrons. Smearing due to zero point
motion is well known. Such an effect features in studies of the $\beta$ spectrum
expected in experiments like KATRIN using diatomic Tritium. 
The recent preprint may, however, challenge new planned experiments
seeking to discover the Cosmic Neutrino Background (CNB) neutrinos
(and/or other neutrinos of masses smaller than $0.1$ eV)
which plan to use Tritium adsorbed onto Graphene or other materials.
Our paper clarifies these issues and examines the more
general problem of smearing induced by quantum uncertainty. 
We find that the effect of Cheipesh {\it et al.} is reduced considerably. The importance of the chemical evolution of the $^{3}$H atom hosting the Tritium nucleus into a tightly bound neutral $^{3}$He atom is emphasized. We estimate the excess blurring caused by the dense spectrum near the lowest state of
the Graphene or other hosts of the Tritium atom, generated by the
electronic response to the ``sudden" escape of the $\beta$ electron.
Our analysis suggests yet larger effects and difficulties facing many
experiments searching for small mass neutrinos. We speculate
on a possible experimental setup which could minimize quantum broadening.

\end{abstract}

\maketitle

 \section{Introduction}
 Smearing due to zero point motion is well known and features in
studies of the $\beta$ spectrum expected in experiments like KATRIN
using Tritium bound in diatomic ($T_2$) molecules \cite{Robertson,thesis}.
 A recent paper by Yevhenia Cheipesh, Vadim Cheianov, and Alexey
Boyarsky \cite{CCB} pointed out that such effects may undermine
Ptolomey type \cite{pt} experiments which plan to use Tritium adsorbed on
Graphene sheets. These experiments \cite{pt} are designed to discover
the Cosmological Neutrino Background (CNB) by careful studies of the
near and beyond the endpoint spectrum of electrons emerging from the
$\beta$  decay of Tritium atoms and from the inverse $\beta$  process
where a CNB neutrino is captured by the Tritium.

An analysis of the CMB data of the Planck collaboration along with
other cosmological data suggested  a most stringent bound on the sum
of the masses of the three neutrinos \cite{Pl2018},
\begin{eqnarray}
\label{sumn}
 \sum_i m(\nu_i) \le 0.12~ \mbox{eV},
 \end{eqnarray}
suggesting that the mass of the lightest neutrino of interest is less than 40 meV.

The lofty goal of CNB discovery faces immense challenges. Detailed 
calculations \cite {CNB capture rates} indicate that, even in
large-scale experiments, only a handful
of the $\beta$ electrons would land in the region of interest beyond
the endpoint of the continuous spectrum. These electrons are
displaced relative to the $Q \simeq 18.6$ KeV endpoint of the continuum spectrum of the
decay electrons by twice the mass of the
background neutrino. If this shift is by less than $0.1$ eV then a
 very high energy resolution will be needed in order to separate the electrons from
the CNB capture from those in the much higher, steeply falling,
continuous Tritium $\beta$ decay spectrum.

\begin{figure}[ht]
\scalebox{.95}{\input{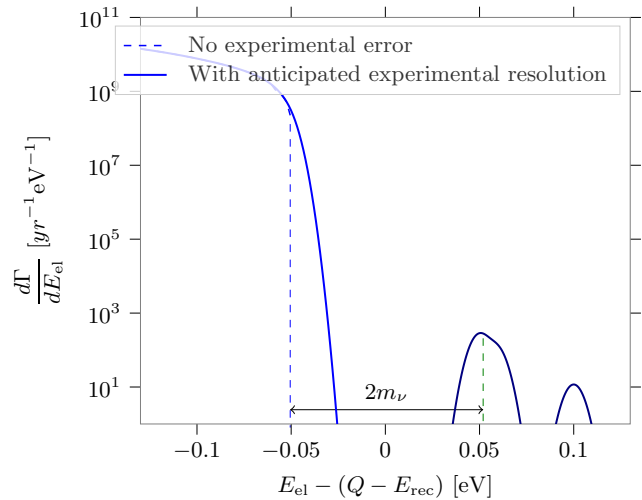}}
\caption{(Color online.) Reproduced from \cite{CCB}. 
The energy spectrum of Tritium $\beta$ decay in the
endpoint region and the peaks due to CNB captures beyond it. Here,
the two CNB masses are 
$m(2) = 2m(1) =0.1$ eV. In the figure, for emphasis, the separation of the peaks exceeds
$\delta(m^2)/(m(2)+m(1))$ (whereas it should be identically equal to the latter ratio). The square difference $\delta(m^2) \equiv
m^2(2)-m^2(1)$ was measured to be
$\sim 8.6 \times 10^{-5}$ eV$^2$. For illustrative purposes, the area under the first peak, proportional to  $|U_{e1}|^2$ (the mixing of $\nu_e$ and $\nu_1$), was exaggerated relative to the area under the second peak (proportional to $|U_{e2}|^2$).}\label{fig:spect}
\end{figure}

The captivating concept underlying these experiments is that in order
to discover the CNB we do {\it{not}} require the precise values of the
expected and/or measured energies at and above the endpoint of the
Tritium $\beta$ decay. Rather, all that is needed is that the measured distribution
$F(\epsilon(\beta(e)))$ of the energy of the $\beta$ electrons will 
display the pattern shown in Ref. \cite{CCB}, reproduced in Fig. \ref{fig:spect}, with approximately a dozen
electrons in a peak (or two peaks) above the observed endpoint. This
holds for {\it any} effect shifting the energy $\epsilon(e)$ in the
endpoint region so long as the shift has a {\it{fixed}}, sharp,
value. This also applies for ongoing experiments like KATRIN searching
for small distortions of the continuum spectrum near the endpoint due
to small neutrino masses. In assessing the prospects of such experiments for detecting CNB we need then
to worry only about variations leading to the blurring of the energy
distribution of the $\beta$ electrons around the above noted $Q \simeq 18.6$ KeV endpoint.

Various ``technical" sources of broadening can be handled. The binding energy
($B$) of the adsorbed $^{3}$H atom of  $\sim$ 1.85 eV exceeds the
resolution needed to establish the CNB induced peak(s) and variations
of the bindings of different Tritium atoms would be harmful. However, this problem may be avoided. 
For instance, in pristine Graphene patches, all binding sites are identical with each
$^3$H atom being equally bound \cite{comment4}. In this situation, the uniformity
disappears at times (as measured from the start of the
experiments) approaching the half life of Tritium ($\tau \sim {13} $
years). At such times, we may find different local arrangements of surviving Tritium atoms.
The random set of Tritium atoms which has decayed is transmuted into Helium ions or
helium atoms which often escape the Graphene sheet. The fact that neighboring Tritium atoms may spontaneously
tunnel and rearrange into a diatomic $T_2$ molecule and evaporate limits on the allowed density of packing. Thermal Doppler broadening in this and gaseous setups can be
minimized by cooling. Finally, the standard energy losses by
ionization of the $\beta$ electrons as they traverse the O(Angstrom)
Graphene sheet are negligible, and on average only one in
$\sim{10^{6}}~ \beta$ electrons will interact on its ``way out'' and lose
energy by ionization. 

In the upcoming discussion, we will employ upper (and lower) case letters to denote quantities related to the heavy $^{3}$H, $^{3}$He$^+$, and $^{3}$He atoms (and with the light
$\beta$ electron and $\nu$) respectively. We will often implicitly set $c$ to unity. However, in order to highlight zero point quantum effects, which form a focus of our work, we will keep factors of $\hbar$. 

The authors of \cite{CCB} recall the theory underscoring irreducible ``noise" due to zero point motion. Their argument is
presented in an idealized context where the hosting Graphene merely
provides a static binding potential which, near its minimum, is
approximated by a harmonic oscillator potential. The Heisenberg
uncertainty principle implies, in the harmonic approximation near the minimum, a momentum spread $\hbar \vec{K}$ of the Tritium atom that is set by the reciprocal of its localization length,
 \begin{equation}
 \label{2}
 \Delta K _i\sim \frac{1}{ 2 \Delta R_i}.
 \end{equation}
Here, $i$ labels Cartesian components of the momentum
and position vectors. 
In the (Gaussian) harmonic oscillator ground state, the more general Heisenberg uncertainty relation inequality is saturated with the above factor of $1/2$ in Eq. (\ref{2}). In this instance, the approximate equality in Eq. (\ref{2}) is precise and estimates made using this relation become exact. The chemisorption potential responsible for the deeper binding of the Tritium to the Graphene is most constraining in the $Z$ direction (with the $Z$ direction defined to be perpendicular to the $XY$ plane of the Graphene sheet). The average (as computed by ab-initio calculations of various groups \cite{A,B,C}) of the rigidity 
 $\kappa$ of the effective harmonic potential near
the minimum of the actual potential then yields an estimate of the amplitude of the vertical oscillations of $\Delta Z \sim 0.13$ Angstrom. Due to the relatively large $^{3}$H mass, this spatial fluctuation scale is much smaller than the length of
the CH bond- the true distance between the $^{3}$H and the Carbon
atom just below (or above) it, of  $\sim{1.2}$ Angstrom. The harmonic
approximation holds over this $0.13$ Angstrom interval. Employing Eq. (\ref{2}), this localization yields $ \hbar c \Delta K_Z \sim 7.7 $ KeV.  With this $K_Z$, the final $^{3}$He$^+$ ion and electron are emitted from a
source moving with a spread of velocities,
\begin{eqnarray}
\label{3}
 \Delta  V_Z = \frac{\hbar \Delta K_Z}{M(^{3}H)} \sim 2.5 \times10^{-6}c.
 \end{eqnarray}
A shallow ``migration" potential limits the lateral motion
along the X and Y directions. The coefficients of the restoring
forces $\kappa_X =\kappa_Y$ are an order of magnitude smaller than
$\kappa_Z$. The resulting standard deviations of the velocity components of the Tritium $\Delta V_X =\Delta V_Y \sim  {\kappa}^{1/4} ~\Delta V_Z \sim 10^{-1/4}  ~\Delta V_Z $. 
The varying boosts then smear the energy distribution $F(\epsilon(\beta(e)))$ of the emitted $\beta$
electron by
 \begin{eqnarray}
 \label{evp}
 \delta \epsilon(\beta (e))  && = \Delta \vec{V} \cdot \vec{p}(e)  \nonumber
 \\ && = \Delta V_X  p_x(e) + \Delta V_Y  p_y(e) + \Delta V_Z  p_x(z) \nonumber
 \\&& =  p(e) \Big( n_x \Delta V_X  + n_y \Delta V_Y + n_z  \Delta V_Z \Big),
 \end{eqnarray}
 where $n_x, n_y$, and $n_z$ are the direction cosines of the electron momentum. The magnitude of the momentum of the electron $|\vec{p}(e)|$ at (and near) the endpoint is 
 \begin{eqnarray}
 p(e) = [2m(e) Q]^{1/2}  \sim 137 ~~\mbox{KeV},
\end{eqnarray}
where we inserted $Q \sim 18.6$ KeV for the endpoint of the $\beta$ electron energy distribution. In the following, we will denote the angular averages by $\langle-\rangle$.  Squaring Eq. (\ref{evp}) and omitting the mixed Cartesian component terms (since those average to zero),
\begin{eqnarray}
\label{triv}
\sqrt{\langle (\delta \epsilon(\beta(e)))^2 \rangle} &&= p(e) \Delta V_{Z}\sqrt{(n_x^2 + n_y^2) (10^{-1/4})^2 + n^2_z} \nonumber
\\ && \sim 0.544~  p(e) \Delta V_{Z} = 0.186 ~\mbox{eV}.
\end{eqnarray}
In Eq. (\ref{triv}), we used $n_x^2+n_y^2+n_z^2=1$ and $\langle n^2_{z} \rangle = 1/3$ and, in the last equality, we substituted the standard deviation of the Tritium $Z$ velocity component of Eq. (\ref{3}). 
This final result for the ZPM induced blurring is significantly lower than that suggested by \cite{CCB}.  The reasons for this discrepancy are twofold:
(a) a factor of 1/2 due to the saturation of the Heisenberg inequality omitted in \cite{CCB} and (b) angular averaging over the uniformly distributed directions of $\vec{p}(e)$. 
Nonetheless, the reduced smearing of Eq. (\ref{triv}) {\it still exceeds} the separation between the
endpoint of the spectrum and the putative tiny peak(s) due to the CNB
neutrino absorption if the mass of the latter is smaller than $0.1$ eV.

Blurring effects of the $\beta$ electron are
associated with a dense spectrum of energies of the system hosting the
initial Tritium nucleus near its lowest energy state induced by  the
escape of the $\beta$ electron. Only final states where the initial $^{3}$H
atom converted into the tightly bound neutral $^{3}$He {\it{atom}} are
associated with the sought, most energetic, $\beta$ electrons.

In the current work, we will discuss, in some detail, three possible settings. 
Specifically, we will examine what transpires when the host system is (a) the initial $^{3}$H atom,
 (b) an $^{3}$H-$^{3}$H molecule for a Tritium source which is molecular or 
 (c) the far more complex situation in which the $^{3}$H atom is adsorbed on a Graphene sheet. 
 Our general framework of a ``dense host energy spectrum" includes the
smearing due to the ZPM where the relevant degree of freedom 
is the linear motion of the $^{3}$He$^+$ ion or the $^{3}$H atom
generated in the $\beta$ process. With only a single degree of freedom involved,
this can be treated exactly. 

Rather than merely viewing the Graphene as the source of the static
potential binding the Tritium, we explore the dynamics of the
Graphene. This involves both phonon (i.e., normal mode vibrations of the Carbon ions in
the two dimensional Graphene lattice) and electronic excitations. Blurring arising from phonon type oscillations is discussed in Section \ref{phonon-sec} as it ties with smearing triggered by the Tritium ZPM that is elaborated in Section \ref{sec:kin}. We find a small excess smearing (by $\sim {12\%}$) due to this effect. By contrast,
the estimates of broadening due to electronic
excitations suggest large effects. Thus, these excitations constitute a formidable barrier for experiments using
Tritium attached to Graphene or other surfaces to discover the CNB
neutrinos. Finally, we briefly suggest a highly speculative scheme of
attaching Tritium atoms to a surface which may reduce broadening
effects due to quantum uncertainties. 

  \section{Kinematics of the CNB capture and smearing by boosts due to
zero point fluctuations}
\label{sec:kin}

 The CNB neutrino in the capture reaction
 \begin{eqnarray}
 \nu(e)+~ ^{3}H  ~\rightarrow e^- ~+~ ^{3}He^+           ,
 \end{eqnarray}
 and the anti-neutrino emitted in the $\beta$ decay process
\begin{eqnarray}
 ^{3}H  \rightarrow e^- ~+~ ^{3}He^+~ +~ \bar{\nu}(e),
 \end{eqnarray}
are eigenstates of electron flavor $\nu(e)$. In the ``normal
hierarchy" of neutrino masses, $\nu(e)$ mixes mainly with the lower
mass eigenstates $\nu(1)$ and $\nu(2)$ which have a small, measured,
$m^2$ splitting \cite{Parameters},

\begin{equation}
 \delta(m^2)_{1,2}= [m(2)]^2 - [m(1)]^2 \sim 8.6 \times 10^{-5}~~ \mbox{(eV)}^2.
 \end{equation}

Detecting two peaks beyond the endpoint, at locations and relative
strengths fixed by $\delta(m^2)$ and by the measured mixings
respectively, may be invaluable for confirming a discovery of the
CNB. The mass $m(1)$ of the lightest neutrino is unknown.  Assuming
$m(1)\sim{m(2)-m(1)}$ and that $m(2) \sim 2m(1)\sim 50$ meV leads to a
$\beta$ spectrum from Tritium decays and CNB neutrino capture reactions as given in 
Fig. 1 in  Ref. \cite {CCB} (reproduced in Fig. \ref{fig:spect}). For higher $m(1)$ values, the need to keep
$\delta(m^2)$ fixed, forces $(m(2)-m(1))$ to decrease and the two
peaks merge at $\sim{2m(1)}$ above the endpoint. In the following,  
we therefore simplify the analysis by considering only {\it{one}} effective light
neutrino $\nu^(1)$ (instead of two) with a large mixing $|U_{e1}|^2 + |U_{e2}|^2 \sim$ 0.95 (close to the maximal value of one (for the full unitary neutrino mixing matrix $U$)) with only $\nu_1$ and its associated single peak (instead of two peaks) above the endpoint.

The kinematics of the $2 \to 2$ body CNB capture process is greatly
simplified by having the $^{3}$H  essentially at rest and the incoming
 non-relativistic CNB neutrino carry almost zero momentum. The
neutrino then just adds its rest mass $m(1)$ to that of the Tritium to
form an intermediate state I at rest in the laboratory frame of mass,
\begin{equation}
      M(I)= M(^{3}H) +m(1).
      \end{equation}
The weak interactions transform this state into the final electron and
the $^{3}H^+$ ion whose center of mass frame is also at rest relative
to the laboratory frame \cite{Fiertz}. The outgoing $\beta$ electron
and $^{3}$He$^+$ ion are then ejected in opposite directions with momenta of
equal magnitude,
\begin{equation}
 p(\beta(e)) = p=P(^{3}H^+) = P.
 \end{equation}
 Energy conservation implies that
\begin{eqnarray}
 \epsilon (e) + E(^{3}He^+)=[ M(^{3}H) +m(1)+ E(^3H) ] \nonumber
 \\  -M(^{3}He^+) -
m(e)+\delta E_{atomic},
\end{eqnarray}
where $E(^{3}H), E(^{3}He^{+}), $ and $\epsilon(e)$ refer to {\it{kinetic}}
energies and $\delta E_{atomic }$  is the change of bindings of the
atomic electron due to the transition from the initial $Z=1$  Hydrogen
to the final $Z=2$ Hydrogen-like Helium ion. The on-shell conditions
for the final electron and Helium ion
\begin{eqnarray}
 p(e)= ((m(e) +\epsilon(e))^2 - m(e)^2)^{1/2}, \nonumber
 \\ P(^{3}H^+) =[(M_{^{3}He^+} +E_{^{3}He^+})^2 - (M_{^{3}He^+})^2]^{1/2},
 \end{eqnarray}
then prescribe sharp values of the kinetic energies of the outgoing
electron and $^{3}$He$^+$ Helium ion. Namely,
 \begin{eqnarray}
 \epsilon(e) +m(e)  = \nonumber
 \\ \frac{[M(^{3}H) +m(1)]^2+[M(^{3}He^+)]^2  -[m(e)]^2}{2(M(^{3}H) +m(1))}  \nonumber
 \\  +\delta E_{atomic},
\end{eqnarray}
and
 \begin{eqnarray}
 E(^{3}He^+)+ M(^{3}He^+)   = \nonumber
 \\  \frac{[M(^{3}H) +m(1)]^2-[M(^{3}He^+)]^2  -[m(e)]^2} {2(M(^{3}H) +m(1))}.
\end{eqnarray}

The above equations express, in terms of the initial kinetic energies, the
familiar relativistic kinematics of the decay of a particle of mass $M$ into
particles of masses $M(1),m(2)$ with {\it{total}} energies
$E(M(1))=\frac{M^2+M^2(1)-m^2(2)}{2M}$ and $\epsilon(m(2))=
\frac{M^2-M^2(1)+m^2(2)}{2M}$.

If the Tritium $^{3}$H atom interacting with the CNB neutrino is bound to, say, a Graphene sheet, then the energy conservation relation will be modified to
\begin{eqnarray}
&&  \epsilon(e)  +E[^3H^+]  \nonumber
 \\ && =[ M(^{3}H) -B +m(1)+ \epsilon(\nu(1))] \nonumber
 \\  && ~~ -M(^{3}He^+) - m(e) +    \delta E_{atomic},
  \end{eqnarray}
 where $B$ is the binding of $^{3}$H to the Graphene. This binding is
smaller than the recoil kinetic energy of $\sim {3.5}$ eV of the
final $^{3}$He$^+$ in the free decay, and we first discuss the case
where the $^{3}$He$^+$ escapes the Graphene. We will first also follow
 \cite{CCB} where all the effects of the host Graphene were subsumed
by a static potential ${\cal{V}}$(Graphene $^{3}H) ={\cal{V}}$.

Unlike the CNB neutrino capture on free $^{3}$H, the capture here is
on a Tritium atom which is bound in a potential and can exchange
momentum $\hbar \vec{K}$ with it. The system of $I=CNB(\nu)+~^{3}H$
and as well as the center mass frame of the final $\beta$ electron and
Helium ion are then boosted relative to the lab frame where
 the energy $\epsilon(\beta(e))$ of the $\beta$ electron is measured.
The resulting change of the energy of the outgoing $\beta$ electron
depends on the magnitude and direction of $\vec{K}$ relative to the
momentum of the electron. Thus, a broadening of the distribution
$F(\epsilon (\beta(e)))$ of the final $\beta$ electron energies is
expected from pure kinematic arguments.
 
 In order to further assess the magnitude of the effect, dynamical information
is required. Before the $\beta$ process happens, say at $t=0$, the
$^{3}$H atom is in the lowest bound state in the
potential ${\cal{V}}$. In momentum space, we will represent the ground state of the
$^{3}$H atom by a wave function $\tilde{\psi}_0(\vec{K})$ - the Fourier transform of the ground wave function in configuration space. The associated momentum probability distribution is given by
$F(\vec{K}) =|\tilde{\psi}_{0} (\vec{K})|^2$.
With the $^{3}$H atom moving in its bound state with momenta $\vec{K}$
distributed according to $F(\vec{K})$ and corresponding velocities
$\vec{V}= \hbar \vec{K}/M_{^{3}H}$, the boosted initial system I of the Tritium
$^{3}H$ with the CNB $\nu(1)$ resting on it, will decay into the final
electron and $^{3}$H$^+$ with each particle having the claimed spread
of energies of $\pm   \hbar \vec{p} \cdot \vec{K} /M$. Viewing the decaying Tritium as a freely evolving wave packet, which
seems to be implied above, would entail energy non-conservation. Each
plane wave component of $\tilde{\psi}_0$ carries a kinetic energy $\hbar^2 K^2/(2M_{^{3}H})$ adding up to the positive
\begin{eqnarray}
 \int d^3K ~  F(\vec{K}) \frac{\hbar^2 \vec{K}^2}{2M_{^{3}H}}.
\end{eqnarray}
Thus, both the final $e^-$ and the $^{3}$He$^+$ ion will have not only
opposite energy shifts $\pm\vec{V} \cdot \vec{p} (e)$ relative to the situation of
free atomic decay. Rather, they will appear to jointly have an extra
kinetic energy not accounted for by the mass differences of
the initial and final particles. This, however, is not the case. The
Tritium atoms with higher kinetic energies are, in a classical
picture, located nearer to the minimum of, and deeper in, the binding
potential. This attractive potential energy is to be subtracted from
that of the outgoing $^{3}$He$^+$ as it leaves the Graphene sheet.
If the potentials exerted by the Graphene on the initial $^{3}$H atom
and the final $^{3}$He$^+$ ion are the same,
\begin{eqnarray}
 {\cal{V}}(^{3}H- Graphene) = {\cal{V}}'(^{3}He^+  - Graphene),
 \end{eqnarray}
then the net effect, also in the quantum case, amounts just to the
binding energy $(-B)$ term in the energy conservation relation above. 

It is instructive to rederive the smearing due to ZPM in a slightly
different way. The oscillation frequencies 
$\omega_i=(\kappa_i/M)^{1/2}$ and the zero mode energies are  $
\hbar\omega_i/2$. In the harmonic oscillator eigenstates, the average kinetic energy is half
of the total energy. Since the expectation of velocity components in
stationary states vanish, the variance 
\begin{eqnarray}
(\Delta V_i)^2 \equiv \langle V_i^2 \rangle  - \langle V_i \rangle ^2 = \langle V_i^2 \rangle =\frac{\hbar (\kappa_i)^{1/2}}{2 M^{3/2}}.
 \end{eqnarray}

The contribution of the zero point oscillations in
the Z direction perpendicular to the Graphene sheet and those of the
X and Y independent oscillations in the in plane simply add up,
reflecting the separability of the harmonic oscillator problem in
Cartesian coordinates. With the average of the $\kappa_3$ values for the chemisorption
minimum computed by the three different groups of \cite{A,B,C} being
$\kappa_3 \sim$ 3.5  eV/Angstrom$^2$, we find that $\hbar
\omega_Z\sim$ 0.07 eV. This recovers the earlier result of $ \Delta V_Z
\sim{2.5 \times 10^{-6}} $ c and the earlier derivation of the ZPM broadening follows again. 

\section{Phonon Excitations}
\label{phonon-sec}

We next proceed to discuss the effect of zero point oscillations of
the Carbon atoms in the Graphene lattice. While the resulting effect
is rather small, its derivation is, nonetheless, quite instructive.
 To estimate the effect we use the Debye temperature. Graphene features notably different sound velocities (corresponding to vibrations in the Graphene (XY) plane and in the Z direction) \cite{sound-Graphene}.
 These phonon spectra are associated with two Debye Temperatures, $T_{Debye}^{(1)}= 2312$
Kelvin and $T_{Debye}^{(2)}= 1287$ Kelvin. The lower energy phonons (corresponding to the Debye temperature of $T_{Debye}^{(2)}$), associated with oscillations in the $Z$ direction, will form the focus of attention next. 
In a circular (or ``2D spherical'') Brillouin zone approximation (similar to that typically employed in the Debye model) for the 2D Graphene sheet, the average phonon frequency (over all modes $\vec{k}$) associated with the Z
direction,
\begin{eqnarray}
\hbar \omega_{Average} \sim \frac{2}{3}  k_{B} T_{Debye}^{(2)} \sim 0.074~ ~\mbox{eV}.
\end{eqnarray}
The two other phonon modes jointly carry almost four times as much energy as the Z mode.
We note, however, that the $^{3}$H atom is at Z $ \sim {\pm~
1.2}$ Angstrom relative to the Graphene sheet. Thus, small amplitude X and Y oscillations of the Carbon
atom that is connected to the Tritium only slightly perturb the $1.2$ Angstrom long
C-H bond that is oriented, in equilibrium, along the Z axis. Conservatively, we will ignore these in plane oscillations.   

In the harmonic approximation to the C-H bond about its minimum, the ZPM of the Carbon atom lying at the other end of ``the spring'' that attached to the $^{3}$H atom is formed by a superposition of many frequencies. The impurity Tritium atom may further give rise to localized phonon modes. Oscillatory perturbing forces acting on the $^{3}$H atom that are associated with random relative phases will add up in quadrature. The four times heavier Carbon implies that the average squared vertical velocities $\langle V_Z^2 \rangle$ of the carbon atoms is only $1/4$ of the average squared vertical velocities $\langle V_Z^2 \rangle$ of the Tritium relative to the bonding carbon. With the velocities adding in quadrature, we therefore expect that the
extra $\Delta V_z$ and associated blurring by $\Delta E$ will
increase only by $\sqrt{1+(1/4)} -1 \sim 11.8\%$ relative to the ZPM effect obtained in the earlier Sections without considering the phonons. 

\section{Two examples}

We next recall two examples from high energy physics of broadening due
to ``zero point" internal motion of electrons in atoms and of nucleons
in the nucleus.
\bigskip

{\underline{{\it Glashow Process.}}} 
The first of these two examples is the Glashow process where a $\bar{\nu}_e$, most likely of
extra-galactic origin, of an extremely large energy,
\begin{eqnarray}
 E(\bar\nu_e)= E(Res) =\frac{M(W)^2}{2m(e)}\sim 6.4 \times 10^{15}~~ \mbox{eV},
 \end{eqnarray}
interacts with terrestrial  atomic electrons producing the mediator
of weak interactions, the W boson, of mass $M(W)\sim {80}$ GeV and
width $\Gamma(W)\sim{2}$  GeV. The rate of the
production of  $W^-$ in the resonant region where
\begin{eqnarray}
  E(\bar{\nu}_e) = E(Res) \Big(1 \pm \frac{\Gamma(W)}{M(W)} \Big),
  \end{eqnarray}
and even in a much broader region, is very large, exceeding the
reaction rate of the other five neutrino species combined. Thus
despite the paucity of such high energy neutrinos some events in the
large Ice-Cube detector were expected and most recently a clear cut
event has been detected \cite{ice}. Glashow \cite{Glashow} noted that ZPM of atomic
electrons broadens the resonance by as much as a factor of two. This
has the amusing and potentially observable consequence of doubling
the $\sim{10}$ Km mean free path of Glashow neutrinos in the earth 
\cite{Lowe}. The few eV kinetic energy of the outgoing $^{3}$He$^+$ in
Tritium decays is  $ \sim {10^{-15}}$ times smaller than the
neutrino energy in the Glashow process, indicating a remarkably wide
range over which the smearing via zero point motion applies.

\bigskip

{\underline{{\it Nuclear Targets.}}}
The second, better known, example involves nuclear targets. The
internal Fermi motion of the nucleons in the nucleus is enhanced by
the exclusion principle and the velocities of these nucleons can reach
$\sim 0.3c$. The resulting shifts of invariant masses of a high energy
projectile and such a nucleon can lower by $30\%$ the minimal energy
required for producing a particle $X$, or a pair $\bar{X}-X$,
relative to the threshold for the same reaction on a free nucleon
target. An analog of this in the present case is that, when the ZPM aligns with $\vec{p}(e)$, the momentum  of the emitted $\beta$ electron the latter has an energy exceeding the ``kinematic bound" of $Q+ 2 m(\nu)$. 
Unfortunately, this ``good effect" is offset and
reversed by the upward shift of the energies of the many electrons
in the steeply falling continuum spectrum just below $Q$.
\bigskip

 There is a qualitative difference between the above two high energy reactions
where both incoming and outgoing particles are (ultra)relativistic
and the present case where the outgoing $^{3}$He$^+$ or $^{3}$He atom
recoils with a velocity: $V=p(\beta(el))/ M(^{3}$He) of  $\sim 3 \times
10^{-5} c$. One may then wonder if, for a Graphene substrate, we
have to account for the fact that the new potential(s)
${\cal{V}}'$ on the $^{3}$He$^+$ ion (or the $^{3}$He atom) may slow down or
even stop their exit from the Graphene sheet an issue that we will
address in detail later. This highlights the fact that, thus far, only
the lowest bound state in the initial potential
${\cal{V}}={\cal{V}}(^{3}$H-Graphene) was used with no direct
reference to this or the new potential
${\cal{V}}'={\cal{V}}'(^{3}$He$^+$-Graphene) following the $\beta$ process (or the one between the
Graphene and a final Helium atom). A main objective in this work is
to rectify this shortcoming and investigate the effect of the
Graphene dynamics- specifically of its electronic excitations- on
the blurring. Indeed, the electrons in the Graphene moving with
velocities that at at the top of the valence band reach a (Fermi) velocity $v_F=
3 \times 10^{-3} c $ generate the potentials between the initial Tritium or
its Helium ion/atom predecessors and the Graphene.

We next turn to a main motif of this paper- that of viewing all smearing effects
as due to the sudden change of the Hamiltonian describing the host
system.

 \section{Smearing via Sudden Changes of the Host System}

Much insight into numerous problems is provided in situations when a
hierarchy of time-scales exists. The shortest here is the time it
takes the $\beta$ electron to escape from the residual $^{3}$He$^+$:

\begin{eqnarray}
 \delta_{\beta-escape} (t) \sim \frac{a_{Bohr}}{v(\beta(e))} \sim
10^{-18}~ sec.
  \end{eqnarray}

In the above, $v(\beta(e))$ is the velocity of the $\beta$ electrons of
energy close to  the endpoint $\epsilon(e)\sim Q = 18.6$ KeV.\begin{eqnarray}
 v(\beta (e)) = \frac{p(e)}{m(e)}= [2Q/m(e)]^{1/2} \sim 0.26 ~c,
 \end{eqnarray}
and $a_{Bohr}\sim 0.55$  Angstrom is the Bohr radius. The velocity
$v(\beta(e))$ exceeds the relevant  velocities of the Hydrogen atomic
electron $\sim{c/{137}}$  by a factor of $\sim 0.26 \times {137} \sim
{35}$.

Thus, the escape of the $\beta$ electron causes a sudden change of the
Hamiltonian describing the various hosts of the decaying Tritium
nucleus. As we discussed earlier, these hosts may be (a) atomic, (b) molecular if the original $^{3}$H is
a member of a $^{3}$H-$^{3}$H molecule, or (c) a solid as in, e.g.,  the $^{3}$H atom absorbed on
a Graphene sheet.
Specifically, in the latter, the initial Hamiltonian for the host
system with the Tritium atom bound to the Graphene transforms at $t=0$
(within the above short $\delta t $ time interval) to the new
Hamiltonian appropriate for the same Graphene sheet with the Tritium
Z=1 nucleus replaced by the Helium Z=2 nucleus at the same initial
location.

In the sudden approximation which applies for all the above hosts,
the time dependent Hamiltonian  interpolating between the initial and
final Hamiltonians acts for too short a time and hence does {\it{not}}
change the {\it global} initial wave-function $| \Psi_i \rangle$ of the host system (defined to be the
complete initial system less the escaping $\beta$ electron). The
energy of the host does however change in the sense that
\begin{eqnarray}
\label{YE}
\Delta{E}  = \langle \Psi_i |{{\cal{H}'}}| \Psi_i \rangle -\langle \Psi_i {|\cal{H}}|
\Psi_i \rangle = E' - E_{0}  \neq {0}.
\end{eqnarray}
Here, ${\cal{H}}$ and ${\cal{H}}'$ denote, respectively, the global system Hamiltonians before and after the $\beta$ process (these Hamiltonians contain the potential energy contributions ${\cal{V}}$ and ${\cal{V}}'$ discussed above). In Eq. (\ref{YE}), we used the fact that the initial low temperature state was very close the ground state of the original Hamiltonian ${\cal{H}}$. An energy change of the environment is indeed expected here as in all of the cases with a time dependent potential. Energy can then be exchanged with the external ``agent" which provides
the time dependent potential- the role of which is played here by
the escaping $\beta$ electron- whose energy far exceeds the atomic
scale energies of $\sim{10}$ eV which are relevant to the problem at hand.

In the eigenbasis of the new Hamiltonian, the standard time
evolution of the initial state is  
$|\Psi(t) \rangle= e^{-i{\cal{H}}'t/\hbar} |\Psi_i \rangle=\sum_n e^{-i E'_{n}t/\hbar} |\Psi'_n \rangle \langle
\Psi'_n|\Psi_i \rangle$.
 The probability $ P(i \to f) =  | \langle \Psi_i | \Psi'_f \rangle|^2$ of winding up in any specific final state $|\Psi'_f
\rangle$ satisfies 
unitarity, $\sum_n P_{i\rightarrow n}= \sum_n P_n=1$, and the total energy shift condition
$\sum_n P_n E_n = E_0 + \Delta E$. For times exceeding the escape time of the $\beta$ electron, we may neglect the
interactions between the exiting $\beta$ electron and the remaining
host system and the conservation of the overall energy
 $\epsilon(\beta(e)) + E' _n(Host)$ ``entangles"  the energy of the
escaping $\beta$ electron with the energy of the final state that the
host winds in. At times larger than the inverse of the energy splitting, the
different states tend to ``decohere" and the system ``settles" into one of
the new energy eigenstates (with the emitted $\beta$ electron 
occupying a continuum state of complementary
energy). This decoherence is related to the time-energy uncertainty relation $\Delta E
\times \Delta t \ge \hbar/2$ which implies that at short times,
after the transition to the new Hamiltonian, the energy of the atomic
electron is not well defined. This also holds for the out-going
$\beta$ electron since energy conservation entangles it with the
atomic system so that both have the same spread $\Delta E
=\delta\epsilon(e))$. As time goes on these energies can get sharper
as $ \sim 1/t$ and smaller variations of levels of the host system get
imprinted on the energy distribution $F(\epsilon (\beta(e)))$ of the
final $\beta$ electrons.

While the $\beta$ electrons travel through the spectrometer
towards the detector, many things can happen to the host system.
However, in general, these developments do {\it{not}} affect the energy of
the $\beta$ electrons and can be ignored in estimates of the smearing.
This holds for any process where energy is exchanged between different
parts of the host system leaving its total energy, and hence also
that of the escaping $\beta$ electron, unchanged. An example is afforded by the decay of the excited
$n=2$ final state of the Hydrogen-like Z $=2$ $^{3}$He$^+$ to the
$n=1$ ground state and photons. The key observation is that the
electro-magnetic field with which the $^{3}$He$^+$ ion shares its
energy, makes, together with the $^{3}$He$^+$ ion, the extended
``environment" hosting the initial Tritium nucleus.

Another example where this feature greatly helps estimate the magnitude of the
blurring effect is that the final $^{3}$He$^+$ ion and/or the $^{3}$H atom,
may, in experiments using Graphene, excite various degrees of freedom
in the Graphene lattice. However, both the Graphene and the atom, in
which the initial radioactive nucleus resided, are parts of the
extended environment and energy transfers between the translational
motion of the recoiling ion/atom and the Graphene will- just like
the radiation of photons in the above example, {\it{not}} change the
energy of the environment- nor that of the beta electron.

In the case of molecular Tritium, one can, in principle, directly
measure the various final states of, say, the Helium Hydride ion
$^{3}$H$^+$-$^{3}$He by tagging it in coincidence with the
measured $\beta$ electron. Indeed, the probability of having various
states the Helium Hydride ion was measured by tagging the recoiling
Hydride and the infra-red photon emitted when the Hydride de-excites
from some level to another. Such tagging could help mitigate
difficulties due to smearing. Unfortunately, it is completely
impractical in the experiments of interest. It takes $\sim{10}$
microseconds-millisecond for the radiative de-excitation of the
Helium Hydride ion and/or for the Hydride to travel  to the
detector. To enable tagging there should be no more than {\it{one}}
coincident $\beta$ electron emitted in this time interval. This
allows for, at most, $3 \times 10^{12}$ $\beta$ electrons to be collected in a year.
As may be inferred from Fig. \ref{fig:spect} and the full spectral shape of the Tritium $\beta$ decay, this is far less than what is required in order to produce the extremely rare CNB capture events. We have no independent information on
the final energy eigenstate $|\Psi'_n \rangle$
that the host winds up in, apart from its energy $E'_n$ (inferred from
$\epsilon(\beta(e))$).

 A key point is that only the cases where after the exit of the
$\beta$ electron the system hosting the Tritium lands in, or very
near to, the {\it{lowest}} possible energy state(s) are relevant. This
is so because by energy conservation the outgoing $\beta$ electrons
then have the {\it{maximal}} energy possible and only such events can
contribute to searches of light mass and/or the CNB neutrinos by
investigating the spectrum of the $\beta$ electrons at and above the
endpoint.

An immediate consequence of all of the above is that we should consider only the cases
when the $^{3}$He$^+$ ion remaining  after the escape of the $\beta$
electron is in its ground state. Also, whenever the tightly bound
$^{3}$He atom can form we should focus on the ``branch" where it did
form and is, furthermore, in the atomic ground state. If possible, this
Helium atom should be bound  to the rest of the host- the other
Hydrogen atom in the molecular case or to the embedding Graphene.
Finally, this last bound state should be in, or near, its ground
state.

To have these peak energy electrons clearly stand out and be separated
from the lower energy crowd, the hosts of  the radioactive Tritium
atoms should satisfy the following two conditions:
\bigskip

$\bullet$ {\underline{Condition A}}

There is a substantial overlap of the initial  state of the host and
the lowest energy eigenstate  $ |{\Psi'_o}\rangle $  of the final
system after the emission of the $\beta$ electron. This single state
can be replaced by a group of final states with energies in a $ \Delta
$ neighborhood of the lowest energy $E'_{\min}$ which jointly have a
large overlap with the initial state of the host.
Specifically the sum over these final states 
$\sum_n {| \langle\Psi'_n|\Psi_i \rangle }|^2$
 should constitute an appreciable fraction of the complete sum over all states
which by unitarity is equal to one. This ensures that in a substantial
fraction of all events, the host winds up with energies
$(E'_{\min}+\Delta) \ge {E'_n} \ge {E'_{\min}} $. \newline

\bigskip

To avoid the extra quantum mechanical smearing from exceeding the experimental
resolution in the measurements of the $\beta$ electron energies
$\delta(\epsilon)_{exp}$, we take $\Delta $ to be equal to the
latter. The CNB peaks generated by electrons associated with the above states,
could still be ``swamped" by neighboring, slightly lower energy,
electrons associated with other final states lying above but close to
$E_{min}$ which collectively form with probability higher than that of
the above ``good" group of states in Condition A. To avoid this we need to satisfy a second condition. \newline

\bigskip

$\bullet$ {\underline{Condition B:}}

The lowest state (or set of lowest states defined in condition A
above) is well separated from the next higher energy  state (or
set of states), which are generated with a higher probability, by
an energy gap larger than $\sim {4\Delta}$.

\bigskip

We next turn to the well-studied cases of atomic \cite{book} and
molecular hosts \cite{Robertson}. These provide concrete and rather
instructive examples of the above general discussion of smearing
within the overarching framework of spectra of states near the lowest
energy generated by the sudden escape of the $\beta$ electron.

\section{Atomic and Molecular Hosts }

The effect of changing potentials on the atomic electron was referred
to above as the $\delta E_{atomic}$ term. The $Z=1$ Coulomb potential
${\cal{V}}$ in the Tritium becomes, after the $\beta$ electron is
emitted, a new potential ${\cal{V}}'$ -the $Z=2$ Coulomb potential of
the $^{3}$He nucleus. Along with the wavefunction, the kinetic energy
of the electron remains the same but the potential attraction is
doubled so that the overall energy change in this case is $
-e^2/{a_{Bohr}} = -2 ~Ry$. Using the hydrogenic $1$S wave functions with Z$=1$ and Z$=2$ for the
initial and final states,  $\Psi_i$ and $\Psi'_f$ respectively, we find that in the sudden approximation the probability that the host
system will stay in the new ground state is
\begin{eqnarray}
 \Big| \Big \langle \Psi _{Z=1} \Big|\Psi'_{Z=2}\Big\rangle \Big|^2
= \frac{2^9}{3^6}\sim 0.64.
 \end{eqnarray}
The fact that the overlap with the new ground state is a substantial
fraction means that condition A is well satisfied here. The spacing of  $3~Ry \sim 41$ eV between the 1S ground state and the first excited 2S state of $^{3}$He$^+$, (a state which in any event
is produced with less than half the probability of the ground state)
is indeed very large. It is $400$(!) times bigger than $\Delta$
which we take equal to a resolution goal of $0.1$ eV so that condition
B is most amply satisfied.

This seems to make atomic Tritium with {\it{no}} intrinsic spreading,
an ideal candidate for sensitive searches for neutrinos.
Unfortunately, it is very difficult to prevent cold atomic Hydrogen
from converting into its molecular form. We thus next turn to the case when the Tritium is part of a diatomic
($^{3}$H -``H") molecule, where ``H'' denotes a Hydrogen or its D or T isotopes.
 We then need to consider the new molecular states appearing after the
sudden escape of the $\beta$ electron.

A notable difference from the atomic realization above, where only one
atomic electron is inherited from the original $^{3}$H atom,  is that
in the molecular case there are {\it{two}} electrons in the immediate
($\sim$ Angstrom) vicinity of the Helium nucleus. Both electrons can
now wind up in the 1S shell of Helium forming a final neutral $^{3}$He
atom in its ground state. The first ionization energy of He is higher
than that of hydrogen by $\sim 12$ eV. This endows the $\beta$
electrons entangled with the final state containing the Helium atom
with energies which are $\sim12$ eV higher as compared with the atomic
case above where only a  single electron winds up in the 1S state in
$^{3}$He$^+$. Again, only these highest energy $\beta$ electrons
make the highest endpoint and putative CNB peak(s) beyond it thereby
contributing to the experimental search. All other $\beta$ electrons
emitted in conjunction with higher final molecular states and/or
ionized Helium will have lower energies and will not play any role in
Tritium endpoint experiments with sub-eV neutrinos.
The probability that the He atom will form is given by the overlap of
the initial state where the two electrons make the covalent $H-H$ bond and
the ground state of the ground state of para-Helium.
 While the fact that both in the initial and final state the spins of
the two electrons are coupled to a total spin $S=0 $ clearly enhances
the probability P(He)  of the branch where an atomic para-Helium is formed, 
the fact that here a pair of electrons (instead of a single
electron) need to
overlap, suggests that P(He) is smaller than the probability that we
obtain the ground state Helium ion in the atomic case of 0.6.

 The final Helium in its ground state can still bind  with the  ionized ``H"$^+$ namely the proton or the deuteron or tritium nuclei. The polarization by a, point like, charged ``H"$^+$ nucleus generates
an attractive potential which at large distances is \cite{ Feinberg}
\begin{eqnarray}
V=-\frac{\alpha_{em} \alpha_{He}}{{2R^4}},
 \end{eqnarray}
with $\alpha_{He}$ the electric polarizability of Helium (and $\alpha_{em}$ the fine structure constant). With no Pauli exclusion to prevent the bare,
positively charged nucleus from penetrating the electron cloud in the
Helium, a significant attractive potential which  extends to small
distances, is being generated. This potential then tightly binds the
Helium atom and the hydrogen nucleus the Helium Hydride ion:
``H"$^{+3}$ He with binding B $\sim {1.68}$ eV  and average radius of $
\sim 0.8 $ Angstrom \cite{Hydride}.

The maximal reduced mass  $\mu(^{3}$H$^+$-$^{3}$He)$\sim 3M(H)/2$
maximizes the binding and minimizes the recoil energy of the  Hydride
in the case of an initial diatomic Tritium molecule. This elevates the endpoint of the
associated $\beta$ spectrum and the putative CNB capture peaks above
the corresponding endpoints and peaks in the decays in TD or TH
molecules- ameliorating the effect of a possible small admixture of
molecules with the lower mass H isotopes. Choosing  $``H" =T$ is also
preferred experimentally because it doubles the amount of radioactive
Tritium and  we will focus on this case.

Calculations of the collective overlap of the initial, post $\beta$
reaction, state and the  ground or excited states of the final
$^{3}$He-T suggested very high values of P$= 0.93$, \cite{Schwartz}
 These values strongly conflicted  with experimental  measurements
which found a probability of $P\sim{0.5}$- an embarrassing state of affairs which
lasted until 1989 when the important paper \cite{ Robertson} rectified it.
A clear review, of both the experimental and theoretical status, and much
original work is contained in the 2015 PhD thesis of Laura I.
Bodine \cite{thesis}.

The significant overlap with the cluster of states within $\sim 0.1$ eV above the ground state of $^{3}$He-T$^+$  implies that condition A above is satisfied. To check if conditions B also holds, we need to
find the behavior of the $E'_n$ energy levels of the Helium Hydride
ion. A study of these states, far more detailed  than what we
present next, is included in \cite{Robertson}, in the above noted thesis of
Laura Bodine \cite{thesis}, and in many yet more recent theoretical works. The
following somewhat heuristic approach may still be useful. The lowest excitations are rotational and the energy of levels of angular momentum$ l$ are
\begin{eqnarray}
 E_l = \frac{\hbar^2 l (l+1)}{ 2\mu R ^2}\sim l (l+1) (1.2) ~{\mbox{meV}},
\end{eqnarray}
where $\mu=M(T)/2  =3 M(H)/2$ is the reduced mass and a molecular
radius $R$ of $\sim 1 $ Angstrom was used. The maximal $l$, fixed by
the total binding in this simplistic approach, is $l_{max}=L\sim {40}$. Both $l$ and $l_Z$ (the
projection along the $\vec {R_1} -\vec{R_2}$ symmetry axis of the
molecule) depend on the uniformly distributed direction of the
momentum of the $\beta$ electron relative to this axis. For each $l$,
there are $2l+1$ states of varying $l_z$ values. Thus, between $l=0$ and a maximal
$l=L\sim{40} $, there are altogether $\sum_{l=0}^{L} {(2l+1)}= L^2 \sim {1600}$  
rotational states of energies extending up to the binding energy of
the Helium Hydride ion of $1.7$ eV with the same average small
separation of $1$ meV between consecutive levels. Also in a simple
classical picture used above, justified in retrospect for the large $l$
values, the probabilities of having the different levels are the
same.

The above discussion focusing solely on the rotational modes suggests that the $1.7
$ eV interval above the lowest energy final state Hydride is
{\it{uniformly}} populated by a dense set of Helium ion bound states,
so that condition B above may still be satisfied. However at  higher
energies the vibrational excitations and their associated rotation
bands restore the well known general tendency of energy spectra to get
{\it{denser}} at higher energies. This, in turn, leads to a violation of
condition B and to appreciable smearing.

Furthermore, the higher vibrational modes that may nearly tear the molecules apart smoothly
connect the bound Hydride ion and its dissociated form of $^{3}$He and T$^+$. We consider the latter 
next. In such decays to three on-shell particles
\begin{eqnarray}
  \nu(1) + (^{3}H ^{3}H) \rightarrow e^- +~^{3}H^+  ~ +~ ^{3}He,
 \end{eqnarray}
the final energies are not sharp. Momentum conservation implies only
that the momenta of the final particles lie in a plane and the total
kinetic energy $E=  E(^{3}T^+) +E(^{3}He)$ of the recoiling atom and
ion depends on the angles between the momenta in this plane. These
angles vary between different events causing a spread $\Delta E$ of
the energy of the heavy particles. Kinematics alone allows $\Delta E$
to extend over an interval of size
 $\Delta E \sim E(\max)- E(\min)= E_{free}- \frac{E_ {free}}{2}=
\frac{E_{ free}}{2} \sim 1.85 ~~{\mbox{eV}}$.
 Here, $E(\min)$ is the minimal total recoil state of interest, corresponding to
equal momenta of $^{3}$H$^+$ and of $^{3}$He so that the $\beta$
electron effectively recoils against a system of mass $2M= 6M(H)$.
 The dynamics prescribing the probabilities of final states with
different recoil energies is encoded in the  momentum space wave
function of the ground state of the initial T-T molecular analog of
$\tilde{\psi}_0(\vec{K})$ mentioned above in connection with smearing
due to ZPM of the tritium atom relative to
the Graphene. We have noted that this ZPM effect can be viewed as part of the
general ``Host connected broadening"  framework.
This clearly applies to the molecular case. The spreading due to  the
different recoil energies of the $^{3}$He atom- the descendant of the
original Tritium- is in fact the very same spreading induced by the
zero point vibration of this Tritium atom and fully accounted for by
the latter.

The H-H bond is $\sim{2.5}$ times stronger than the  C-H
chemisorption potential used in estimating the zero point smearing
effect for Tritium adsorbed on Graphene and the reduced mass here is
$5/8$ times smaller. This renders the ZPM blurring effect here twice as
large $\sim 0.48$ eV. This, however, may not be extremely important as
(i) these dissociated states lie above the lower energy bound states of the
Helium Hydride ion and (ii) kinematics implies that dissociation
does {\it{not}} occur in most cases. 

In principle, an exhaustive calculation can be done (and in view
of the big experimental effort is indeed ongoing). This requires
detailed computations of the wave functions and energies of the rotational,
vibrational levels of the $^{3}$He-``H"$^+$ molecule-ion, and the
overlaps of these states with the state $\Psi_i$ generated
immediately after the $\beta$ process happened at $t=0$. The
locations $\vec{R}_1$ of the newly formed $^{3}$He and $\vec{R}_2$
of the other ``H" are the same as those of the two atoms in the
$^{3}$H-``H" molecule and the $^{3}$He has the momentum kick $\vec{P}
= -\vec{p}(\beta(e))$. In real space, this wave-function assumes
the form
 \begin{eqnarray}
\langle \Psi_i| \vec{R}_1, \vec{R}_2 \rangle  =
e^{i\vec{P} \cdot \vec{R}_1/\hbar} \Psi_{^3H-``H"}(\vec{R}_1-\vec{R}_2),
 \end{eqnarray}
where $\Psi_{^3H-``H"}$ is the initial wave function of the H-``H"
molecule in its ground state described in terms of the H-``H"  relative
separation
 $\vec{R} =\vec{R}_1-\vec{R}_2$. The factor $\exp[i\vec{P} \cdot \vec{R}_1/\hbar]$
``imparts'' a recoil momentum $\vec{P}$ at the location of the $^{3}$H atom.

Presently, the KATRIN experiment using molecular $T_2$ is  hampered
by experimental difficulties. A recent publication \cite{KATRIN} presenting KARTIN's
latest upper bound on the mass of the lightest neutrino of $1.1$ eV,
listed the many experimental issues that prevented reaching the
initial, much more ambitious, goal of KATRIN.

     \section{Graphene revisited}
     \label{sec:Graphene}

Using Tritium stored inside solid materials and/or adsorbed on
surfaces may reduce the experimental difficulties associated with
monitoring large quantities of gaseous Tritium by using large $\beta$
spectrometers. In a most optimistic scenario, the experimental
technique may improve to the point of allowing searches for light CNB. In this case, a careful study of the irreducible quantum noise in the new proposed experiments is of paramount importance. We continue to do so here for the ``benchmark" case of Tritium adsorbed on a Graphene sheet.

 Our discussion of the molecular Tritium decays showed that adding
 {\it{one}} extra particle to the Tritium atom generated significant
 smearing thanks to the dense spectrum of final $^{3}$He-``H" states. As
 we argue in this section, the much more complex excitations spectrum
 of the many body Graphene system yields much stronger smearing when
 the Tritium is adsorbed onto Graphene.

The smearing effects fall into two categories: (i) blurring due to nuclear motions
 and those due to (ii) electronic excitations. The overall scale of
 electronic energies is higher than that of  the nuclear  vibrational/
 rotational modes. Still, {\it{all}} smearing encountered above were due
 to the dense spectrum of the latter. Thus the large dispersion by more
 than $\sim{40 }$ eV of the energy spectrum of the final Helium ion
 caused no harmful smearing in the case of atomic Tritium decays
 because of the extremely sparse nature of this spectrum. The Graphene
 has a similar large span of a $\sim$15 eV broad valence band. However,
 the energy spectrum in this band is continuous and dense suggesting
 extensive smearing.

 Can we approach this problem in a more quantitative fashion?
 The rapid escape of the $\beta$ electron justifies using the sudden
approximation. In this approximation, we have at $t=\delta t  (\beta$~ escape),
 the Z=2 Helium nucleus, endowed with the free recoil energy sitting
 at  $\vec{R} =(0, 0, Z_0)$ the original location of the parent
 Tritium which is shifted relative to its binding carbon atom at the
 origin by $\sim{Z_0 = 1.2} $ Angstrom.
 To further justify the sudden approximation and better realize the
 relatively large number of electrons involved, we note the following:
 For a distance $\rho$ from the location at $(0,0,Z_0)$ of the initial Tritium, it takes a time of only $\rho/c$ to establish the new potential generated after the escape of the negative $\beta$ electron. 
 This region contains $N_C = \pi (ct)^2 \times $ the number
 density of Carbon atoms in Graphene (of $3.82 \times 10^{15}$
 cm$^{-2}$ atoms). There is only one free electron per Carbon as three
 out of the four outer shell electrons form $sp^2$ hybridized covalent bonds with the three neighboring
  Carbons. Thus, the same number of  $N_e=N_C$ electrons will be
 influenced. The final Helium nucleus moves very slowly with velocity
 $V \sim 3 \times 10^{-5} c$. Using the size of the Graphene unit
 cell, we find that by the time it moved just $\ 0.1$ Angstrom, $N_C
\sim 0.9 \times 10^7$ Carbon atoms and an equal number of 
 electrons are affected. For the purpose of our qualitative discussion,
the number of electrons that can reach the origin in this time
interval may serve as a measure of the complexity actually
encountered. Using electron velocities of the order of the Fermi
velocity $v_{electron} \sim v_F = 3 \times 10^{-3} c$, which is appropriate
near the top of the valence band, we then find that ``only"
 $(v_F/c)^2 N_c \sim 100$ electrons are likely to be engaged.
For a complete calculation we need to find  the probabilities  for
``exciting" (in the sudden approximation) any one  of the many levels
of the Graphene in either one of the two main ``branches" where the
Helium ion $^{3}$He$^+$ or the Helium atom $^{3}$He has formed. In
principle, this requires the full wave-function of the initial ground
state of the Graphene $N$-body system with the bound Tritium impurity
at $\vec{R}=(0,0,Z_0)$. Next we need to compute the wave functions
and energies of the states arising when the Z=1 Tritium nucleus
is replaced, after the escape of the $\beta$ electron, by the Z=2
Helium nucleus. Finally, we must compute the overlaps of the initial
state with each of these many final energy eigenstates so as to
produce the spectrum of excitations. At first sight, such an undertaking seems impossible. While the ground state of perfect Graphene is (almost) fully understood, the study described above requires  extensive
calculations. The simplifying concepts of conduction/valence
bands rely on the invariance with respect to translation by lattice
vectors, an invariance which is maximally broken near the impurities. However, the very fact that the initial and final Hamiltonians $\cal{H}$ and $\cal{H'}$ differ only locally, in the region near
the ``impurities", is of great help. It guarantees that the overall
energy change induced by the sudden $\beta$ process $\Delta E$
will be $\le{10}$ eV and that only a limited number of electrons could
be affected.

We can further take advantage of this locality by using a finite grid to
simulate the two dimensional Graphene lattice around the impurity of
interest. As an example, we recall the well studied case \cite{Lyang}
of the electron-hole exciton in Graphene where such ab-initio
calculations correctly predicted the exciton energy levels and response
to radiation.

Equally important is the fact that in order to estimate the spreading
near the endpoint, we do {\it{not}} need the complete excitation
energy spectrum- but only in the vicinity of the new ground state. As
emphasized, this ground state is  in the ``Helium branch'' of the
final states which contain Helium in its ground state. Hopefully, this
branch has a significant weight as otherwise investigation of the end
point and vicinity will be plagued by smearing {\it{and}} by a reduced
statistics.

The relatively deeply bound (by 1.7 eV) Helium Hydride ion plays an
important role in experiments involving molecular Tritium. Naturally,
one may wonder how strongly bound is the Helium atom to the rest of
the system (that, in the present context, is the entire Graphene sheet).
Note that because the electron mass is $3000 $ times smaller than
the reduced mass in the case of the Helium Hydride, there
is, most likely, {\it{no}} bound state of  Helium and a {\it{free}} electron. However, the
electrons inside the Graphene- and, in the present situation, the hole which
the escape of the $\beta$ electron created, are strongly bound to the
Graphene so as to effectively carry all of its inertia. We will not attempt here to calculate this binding beyond noting that
it is likely to be stronger than that of the Helium Hydride ion. The reduced mass here is $\mu=$ M($^{3}$He)= 3~M(H) instead of $\mu=3
M(H)/2$. The invariance with respect to shifts of the Helium atom
parallel to the (XY) plane of the Graphene sheet suggests that the
problem effectively become one dimensional. The long range $R^{-4}$
attractive potential between the Helium and a point-like charge
\cite{Feinberg} becomes $Z^{-4}$ with the same coefficient and with the kinetic terms associated with momenta in the XY plane omitted, the kinetic energy is further reduced by $\sim {66\%}$.

 Let us assume that future calculations will reveal that  
 (a) The Helium is indeed bound to the Graphene sheet with a very large
binding energy and that as for the Helium Hydride ion in the molecular
Tritium case there is a considerable overlap with this particular state
{\it{yet}}
 (b) unlike for the above Helium Hydride, there are {\it{no}} such
important excited Graphene-Helium states.

In analogy with the discussion of atomic Tritium, we may then expect {\it{no}} smearing. Unfortunately, this is not the situation here as there is a continuum of Graphene electronic excitations.
Specifically, this amounts to saying that assumption (b) is untenable.
Even if the potential model and the motion of the bound Tritium alone
do not lead to a dense excitation spectrum, the electronic excitations
of the Graphene may do so.

We next present a simple estimate of the smearing due to the
 electronic excitations in the Graphene. We saw above that up to 90
 electrons can reach the origin- i.e., the location of the carbon atom
 situated just below the original adsorbed $^3$H atom, while the final
 $^3$He$^+$ or $^{3}$He move a distance of just $0.1$ Angstrom. Only a
 small fraction of these electrons actually ``move" to the immediate
 vicinity of the origin- defined here as the unit
 cell around the binding carbon at  $\vec{R}(t=0) = (0,0,0)$ as
 their migration is blocked by both the Pauli exclusion and by mutual
 Coulomb repulsion. Only one unit of negative charge is needed in order
 to neutralize the positive charge generated by the sudden escape of
 the $\beta$ electron. This single extra electron can affect the
 formation of the neutral final Helium atom- the branch that we focus on.
 As the relevant electrons are initially localized within distances
 of $\Delta r \sim  {1- 3} $ Angstrom, away
 from the $\beta$ decay event, their momentum spread is
   $     \hbar c \Delta k \sim \frac{\hbar c}{2 \Delta r}  \sim  0.33- 1 ~~\mbox{KeV}$.
 Given that $\Delta{r}$ is similar to the size of the lattice unit cell,
 such a momentum spread, close to the energy scale of  the Brillouin zone,
 is indeed expected. Electrons shifted in energy by less than
 $\Delta E'=$ 1 eV from the Dirac point obey a linear dispersion rule $E(k) =\hbar v_{F}
 k$. Substituting the Graphene Fermi velocity $v_F=3 \times10^{-3}  c$ yields a
 spreading of energies of size
 \begin{equation}
 \label{EVV}
 \Delta E \sim \hbar v_{F } \Delta k = \frac{ \hbar v_F}{2 \Delta r} \sim 1-3~~ {\mbox{eV}}.
 \end{equation}
  The above rather heuristic arguments do not replace a direct
 evaluation of the overlaps between the initial state shortly after the
 escape of the $\beta$ electron with the lowest energy cluster of final
 eigenstates of the new Hamiltonian $\cal{ H'}$. However, finding which
 electrons in the Graphene are ``mobilized" by the new Hamiltonian
 helps estimating the expected smearing. This type of information is
 encoded in the time evolution
$ |\Psi(t) \rangle= e^{-i {\cal H}'t/\hbar} |\Psi_i \rangle$ 
 of the initial wave function under the new Hamiltonian when expressed
in configuration space rather than in terms of the new energy
eigenstates. While this does not achieve
the required projection onto the subspace of the lowest energy
states, it gives us an idea of the overall energy shift induced by the $\beta$ process. The claimed smearing
then follows if this total shift is of the order of an eV {\it{and}} no
prominent low energy state can account for this energy change as was
the case for atomic Tritium decays. Finally, the dual use above of both
momentum and coordinate space is naturally called for by the problem
at hand with the extended lattice and its bands on the one hand and the
local impurity at the origin on the other.

 A slightly different, more succinct, way of rephrasing the argument
is the following: The important change instigated by the escape of the $\beta$ electron
is that  the region near to the original Tritium becomes positively
charged. One can therefore expect significant overlap with new energy
eigenstates where negative charges moved towards the origin in order
to neutralize this positive charge. The fact that the holes generated in this process are produced at
various depths in the band means that the Graphene system will have
this spread of energies which, in turn, will be reflected in a 
substantial smearing of the $\beta$ electrons
energy distribution $F(\epsilon(\beta(e)))$. The lowest energy holes are at the Dirac point at the tip of the cone
at the top of the valence band. This single point has zero measure.
However, the area of the intersection of the conical valence band and
fixed energy planes increases quadratically with the (absolute) value
of the energy of the hole like $E(h)^2$  as we go deeper into the band. 
 This, then, causes a smearing which, in principle, could be as large as the
 full width of the valence band of $\sim{ 15}$ eV.

 So far, we have largely ignored the branch with the Helium ion $^{3}$He$^+$ in the final state as the energies in this branch are upward shifted relative to the Helium atom branch by $\sim{12}$ eV. In the
 remainder of this section we will address this important branch. This is
 required in order to gain a better picture of the overall $\beta$ electron
 spectrum, also away from the immediate vicinity of the endpoint.
 A first observation is that the arguments re overlaps of the
 ``initial" state, formed early on, at times $\sim
{\Delta t_{\beta(e) escape}}$ before the branching into the Helium atom or Helium ion sector occurred,
 equally apply to the Helium ion branch. 
 
The momentum of the heavy
descendant of the initial Tritium, be it the Helium ion or the Helium
atom is large, $P(^{3}He^+) \sim P(^3 He) = P  =p(\beta(e))$
are large enough so that $RP \sim 85 ~\hbar$ for $R \sim 1.2 $ Angstrom- the CH bond
length. We can then approximate the motion of the heavy particles on
scales larger than, say, $0.5$ Angstrom as classical and deterministic.
By the late time that this distance is reached, the Born Oppenheimer
approximation may become applicable.\cite{B.O approximation}

The Graphene sheet will then behave as an ideal conductor with the electronic distribution mimicking a
negative image charge that is situated at $(-Z)$. The resulting attractive potential energy ${\cal{V}}'=-e^2/(2Z)$ assumes at  $Z=1.1$ Angstrom $= 2 a_{Bohr}$ the value of  $\sim{0.5}~$Ry $= 6.8$ eV. This potential energy is much
larger, in absolute value, than the initial kinetic energy $\sim {1.85}$ eV (the free recoil energy $\sim{3.7} $ eV minus the binding energy of  B $\sim{1.85}$ eV). As such, electrostatic interactions will reverse the motion of the outgoing Helium ion. We will not follow the full history of the motion of the Helium ion in this one dimensional (solvable) $1/Z$ potential. We must remark, however, that the probability of {\it{eventually}} having a final He atom, is greatly
enhanced beyond the fraction of Helium atoms in the ``Helium branch"
directly generated by large overlaps in the sudden approximation.
This enhancement is due to late captures of a Graphene electron by
the Helium ion upon its multiple re-entries into (or reflections
from) the Graphene sheet. In successive encounters, the Helium ion
also keeps losing its kinetic energy making it eventually stop in a
bound state within the Graphene. 

As we emphasized earlier, such energy exchange will {\it{not}} be reflected in the measured $\beta$
electron spectrum. The late formation of Helium atoms will, furthermore, not
contribute to the ``good"  part of the $\beta$ spectrum associated with
Helium atom  branch. At long times, the 
coherence and the entanglement are lost. These would show up, however, in coincidence experiments which will
indicate a very high percentage of events with Helium atom, rather
than a Helium ion in the final state. In the above scenario,
the Helium ion will survive only in the rare instances where the initial
energy of the ZPM of the Tritium was so large that the Helium ion it
converted into can escape. Similar multiple encounters with late $^{3}$He formation can happen
also in the molecular case thereby enhancing the measured fraction of
Helium atoms and reducing that of Helium ions in the ultimate final
state.

\section{Summary and conclusions}

This work was inspired by the claim of \cite{CCB} that quantum uncertainty poses a fundamental limitation for detecting CNB. Specifically, Ref. \cite{CCB} underscored that zero point
motion (ZPM) induced smearing of the energies of the $\beta$ electrons
 block experiments aiming to discover the CNB made of neutrinos of
masses $\le{ 0.1}$ eV by using Tritium bound to a Graphene or to other
surfaces. If true, this presents a notable obstacle to
experiments that aim to measure the mass of light neutrinos.
Recalculating the effect we found that it is significantly
smaller than the suggestion of \cite{CCB}. It still remains a challenge for the
lighter $\le{0.1}$ eV CNB neutrinos. We supported this effect by
kinematic arguments and by recalling two instances of high energy
scattering in which atomic or nuclear ZPM play important roles. We
also noted that the ZPM due to the phonon degree of freedom of the
Graphene slightly augments the broadening due to the direct ZPM of the
Tritium.

 Much of our effort was directed towards estimating the broadening due
to electronic excitations induced in the Graphene by the ``sudden"
disappearance of a unit of negative electric charge due to the escape
of the $\beta$ electron. This was done along with the simpler situations of
atomic and molecular $^3$H within a unified approach where all
smearing are viewed as being due to excitations (or more general
energy changes) of the system hosting the initial radioactive Tritium.
Our estimates suggest that electronic excitations of the
Graphene generate a much larger smearing of the measured energy than the above
ZPM effects.

 Can we, following a similar suggestion in Ref. \cite{CCB}, ameliorate the smearing by using appropriate materials to which the Tritium may be attached? Platinum readily adsorbs Hydrogen but binds it more weakly than Graphene with a potentially reduced smearing due to the ZPM of the Tritium. Unfortunately, the surfaces of the thin platinum sheets are not a regular as those in Graphene.

In general, it will be difficult to avoid at the same time excessive
smearing due to the ZPM of the Tritium atom {\it{and}} the smearing due
to a dense spectrum of low energy electronic excitations of the host
material. Our quest may thus lead us back to free atomic Tritium
which, as we saw above, is effectively free from smearing.
  To approach a state of affairs emulating atomic Tritium we have to
a) spatially separate the Tritium atoms so that they will not combine
into molecules, and b) The Tritium atoms should be only weakly coupled
to an embedding ``matrix" material used as otherwise the electronic
excitations of the latter could generate excessive smearing.

 The desire to fulfill both demands then suggests the following setup.
 Inspired by the dark matter search experiment using DNA strands
hanging from gold sheets suggested some time ago \cite{Drukier}, we
would like to have the Tritium atoms weakly attached to the ends of
polymer chains hanging from a thin sheet. If neighboring polymer segments attached to the matrix sheet are (evenly)
spaced by a distance $d$ exceeding the common length $l$ of the
segments then spontaneous formation of $T_2$ molecules will be avoided
regardless of how much the segments keep thermaly dangling.
We still may need to verify that the affinity of the Tritium to the
matrix sheet will be minimal (far smaller than that of in the above
discussion of Graphene) in order to avoid the sticking of the tritium ends
to the surface after accidental looping back of the polymer chain so
as to touch this surface. At distances $\ge{ 50}$ Angstrom, electronic excitations in the
matrix sheet will be largely decoupled from the far-away Tritium and a
potentially large resulting smearing- similar to that found
above for Tritium attached directly to a Graphene sheet- will be
avoided.

Finally, if the bond via which the Tritium is attached to the end of
the polymer chain is weak and the corresponding binding potential is
very shallow the ZPM of the tritium may be significantly reduced and
the attendant smearing far less than the 0.24 eV value found above.
None of the dangling and backward looping was a concern in the setup
suggested in \cite{Drukier}. The very long and heavy DNA chains will
{\it{not}} freely dangle. This is because the gravitational energy
in such setups exceeds the thermal energy $k_{B}T$ limiting the amplitude of
upward thermal fluctuations.
An important factor avoiding such upward jumps of the ends of even
short DNA segments as in the present case is the stiffness of the
double helical DNA. With DNA, we can further harness bio-processes (such as PCR)
for mass production of the segments and possibly also for the attachment
of the segments to the matrix sheet \cite{Other isotopes}.
While the suggested setup seems to be rather complicated and difficult to implement, the goal of
discovering the CNB neutrinos and the absolute scale of neutrino masses may well be worth it.

 {\bf{Acknowledgments}} \newline
It is a pleasure to thank Li Yang and Xiaobo Lu for illuminating discussions, careful reading of our work, and detailed comments. We enjoyed the help and encouragement of Robert Shrock and wish to
thank Erik Henriksen for a useful discussion. We further wish to acknowledge the Aspen Center for Physics, which is supported by NSF grant PHY-1607611.

\onecolumngrid


\begin{thebibliography}{28}
\bibitem{comment4} To be precise, this holds true when only a small
fraction of the sites are occupied and also in the opposite situation where
the surface is completely saturated with Tritium.
\bibitem{Fiertz}  The momentum transfers in the CNB capture reaction
are much lower than the  $\sim{100}$ Mev nuclear scale momenta.  This
allows treating the $^{3}$H and $^{3}$He spin half nuclei as point-
like fermions. The  $(V-A)X (V-A)$  local four fermion weak
interaction can be Fierz rearranged into a $(V-A)X(V-A)$ form in the
crossed (s) channel,

\begin{eqnarray}
\bar{\psi}_{^3He^+}(x) \gamma_\mu( 1-\gamma_5) {\psi}_{\nu(e)}(x) \cdot
\bar{\psi}_{^3H}(x)\gamma_{\mu}(1-\gamma_5) \psi_{e}(x).
\end{eqnarray}

This form  clearly exhibits  the transition from the initial state I of
$\nu_e$ and $^{3}$H into the final state of  $^{3}$He$^+$ and $e^-$
\bibitem{Helem}
Conceivably, the electrically conducting DNA features a dense spectrum of electronic excitations that undermine our suggestion. A possible, admittedly rather difficult to implement, solution to the problem of backwards looping and sticking could be the following. If the length of the flexible polymer segment at the end of which the Tritium is located is 
at $\ell \simeq $ 50 Angstrom then to avoiding thermal backward looping, we need to add a mass element $m$ at (or near) the tip such that $mgh \ge k_{B} T$. For $T \sim 1$ Kelvin, a mass of $m \simeq 10^{-12}$ gram which is that of a spherical iron grain of radius $R \simeq 3 \times 10^{-5}$ cm.
\bibitem {Other isotopes}
The possibility of using $\beta$ decaying isotopes other than Tritium
to study neutrinos has been discussed several times. (For a recent
suggestion, see \cite{Smith}.) Isotopes undergoing super-allowed $\beta$ decays like Tritium but with
lower Q value would have an enhanced (by $[\Delta/Q]^3$) fractional
population of the near endpoint region of interest of electrons of
energies $(Q-\Delta)\le \epsilon(\beta(e))\le Q$. The effect of a
fixed neutrino mass would be much larger and $\beta$ spectrometers with weaker electric fields
will be required. Also the ZPM of the heavy Isotope and
attendant smearing decrease like the inverse of the mass of the
isotope. Unfortunately we do not have such isotopes with lower Q values which
are still super-allowed. The well known $^{187} Re$ isotope with
 $Q= 2.5$ KeV has a a forbidden transition with too low decay rates.
\end{thebibliography}
\end{document}